\documentclass{Interspeech_Arxiv}

\interspeechcameraready

\title{Improving Multilingual Speech Models on ML-SUPERB~2.0: Fine-tuning with Data Augmentation and LID-Aware CTC}

\author[equalcontribution]{Qingzheng}{Wang}
\author[equalcontribution]{Jiancheng}{Sun}
\author[]{Yifan}{Peng}
\author[]{Shinji}{Watanabe}

\affiliation[nocounter]{}{Carnegie Mellon University}{USA}
\email{\{qingzhew, jianches, yifanpen\}@andrew.cmu.edu, shinjiw@ieee.org}

\usepackage{comment}
\usepackage{enumitem,amssymb}
\usepackage{multirow}
\usepackage{booktabs}
\usepackage{xcolor}
\usepackage{siunitx}
\usepackage{float}
\usepackage{amsmath}
\usepackage{bold-extra}
\usepackage{svg}
\newlist{todolist}{itemize}{2}

\setlist[todolist]{label=$\square$}
\DeclareSIUnit{\billion}{B}
\DeclareSIUnit{\million}{M}

\begin{document}
\maketitle
\begin{abstract}
Multilingual speech processing with self-supervised or supervised pre-trained Speech Foundation Models (SFM) has achieved strong performance on tasks like Language Identification (LID) and Automatic Speech Recognition (ASR). However, these models struggle with limited resources during fine-tuning. This paper enhances multilingual LID and ASR on ML-SUPERB~2.0 by exploring multiple strategies for adapting SFMs, including frozen upstream training, partial fine-tuning, and low-rank adaptation. Furthermore, we employ data augmentation to mitigate performance gaps in few-shot settings and introduce LID Connectionist Temporal Classification (CTC) loss for regularization. Our approach achieves a 14\% relative improvement in LID accuracy and a 30\% relative reduction in ASR CER over the baseline on ML-SUPERB~2.0, securing second place in the Interspeech 2025 ML-SUPERB~2.0 Challenge.
\end{abstract}

\section{Introduction}

Multilingual speech processing, including tasks such as Language Identification (LID) and Automatic Speech Processing (ASR), is essential in a world with more than 7000 languages \mbox{\cite[p.~1]{austin2011camblang}}. Recent advancements in multilingual speech foundation models (SFM) based on self-supervised learning (SSL) have significantly expanded the scope of robust multilingual speech processing capabilities to at most over 4000 languages with large-scale unlabeled data~\cite{babu2021xls, li2022asr2k, chen2023wavlablm, pratap2024mms, chen2024xeus, boito2024mhubert}.
In parallel, the supervised pre-trained SFM has also advanced, enabling scalable solutions across over 100 languages and multiple tasks~\cite{zhou2021acm, radford2023whisper, peng2024owsm, peng2024owsmctc}. 

To evaluate SFM in multilingual LID and ASR, ML-SUPERB~2.0~\cite{ml-superb2} is introduced as an enhanced benchmark, covering over 140 languages and 56 dialects. 
Unlike ML-SUPERB~\cite{ml-superb}, which restricted evaluation to frozen upstream parameters and fixed architectures, ML-SUPERB~2.0 adopts a more flexible framework without strategy restrictions. 
This flexibility enables broader evaluation across diverse models, strategies, and learning scenarios. 

While various strategies and models have been evaluated on ML-SUPERB~2.0, prior work reveals significant deviations in LID and ASR performance across languages~\cite{ml-superb2}.
Notably, both performances degrade substantially for languages under few-shot training conditions~\cite{ml-superb2}. 
Another key challenge is that fine-tuning on low-resource languages with limited labeled data leads to overfitting and poor generalization to unseen dialects.

In this paper, we propose to improve the performance of multilingual LID and ASR on ML-SUPERB~2.0. 
Specifically, our work focuses on three key aspects: 
First, we investigate training strategies for adapting SFMs on ML-SUPERB~2.0, including frozen upstream training, partial fine-tuning, and Low-Rank Adaptation (LoRA).
Second, to improve performance on low-resource languages, we incorporate additional labeled data into the training data.
Third, to regularize fine-tuning, we employ an auxiliary Connectionist Temporal Classification (CTC)~\cite{graves2006ctc, tjandra2020deja, lee2021intermediate, nozaki2021relaxing, chen2023lidctc} loss for LID prediction during SFM fine-tuning. 
This enhances generalization to unseen dialects. 
Our approach achieves an LID accuracy of 86.9\% and an ASR Character Error Rate (CER) of 15.6\% on ML-SUPERB~2.0.

\section{Related studies}
\label{sec:related studies}

\textbf{Multilingual speech foundation models.} 
Both supervised and self-supervised training strategies are used to build multilingual SFMs. 
Self-supervised pre-trained SFMs like Massively Multilingual Speech (MMS)~\cite{pratap2024mms} and XEUS~\cite{chen2024xeus}, trained on 1400+ and 4000+ languages respectively, achieve remarkable multilingual performance.  
Meanwhile, supervised pre-trained SFMs such as encoder-decoder-based Whisper~\cite{radford2023whisper} (99 languages) and encoder-only OWSM-CTC~\cite{peng2024owsmctc} (151 languages) rely on labeled speech during training.  
In this paper, we conduct evaluations of three representative encoder-only multilingual SFMs (MMS, XEUS, and OWSM-CTC) on ML-SUPERB~2.0.

\textbf{Data augmentation.} 
Prior work has explored enhancing ASR model performance on unknown languages by incorporating additional language data for training~\cite{li2022lifelong}. 
Another work~\cite{barteldsetal2023making} enhances low-resource ASR by leveraging self-training or a text-to-speech system to generate more training data.
Speed perturbation~\cite{ko2015speedperturb} and SpecAugment~\cite{park19e_interspeech_specaug} augment audio or feature inputs to improve the training data quality. 
In contrast, our work focuses on improving underperforming languages utilizing additional labeled data from public datasets.

\textbf{LID-aware training.} 
The intermediate CTC loss has been shown to improve robustness by introducing auxiliary supervision at intermediate layers~\cite{lee2021intermediate}.
To leverage LID supervision, previous work like~\cite{chen2023lidctc} applies an auxiliary LID CTC loss at the \textit{downstream} intermediate encoder layers, conditioning later layers on LID predictions. 
Another approach extracts language-related frames by average pooling to guide language prediction~\cite{xue2024sshr}.
Unlike these approaches, we incorporate LID CTC loss directly into \textit{upstream} fine-tuning, enabling language-aware representation learning earlier in the upstream.

\section{Proposed method}

\subsection{Training strategies with multilingual SFMs}

To evaluate the performance of the latest multilingual SFMs on the ML-SUPERB~2.0 benchmark, we select three representative encoder-only SFMs as the upstream models: the SSL-based MMS~\cite{pratap2024mms} and XEUS~\cite{chen2024xeus}, and supervised pre-trained OWSM-CTC~\cite{peng2024owsmctc}. 
The overall architecture, as shown in Figure~\ref{fig:model}, consists of an SFM as the upstream encoder, followed by a downstream model trained with the CTC loss.  
Three training strategies are adopted: training downstream with upstream frozen, upstream model fine-tuning, and parameter-efficient upstream model adaptation. 

\subsubsection{Upstream models}

MMS and XEUS share a common architecture, both starting with a convolutional waveform encoder that extracts a $T$-length sequence of $D$-dimensional acoustic features $X \in \mathbb{R}^{T \times D}$ from raw audio. 
These features are then processed by a stack of $L$ encoder layers ($\{f_{\text{ENC}}^l\}_{l=1}^L$), using Transformer~\cite{vaswani2017transformer} in MMS and E-Branchformer~\cite{10022656ebranchformer} in XEUS: 
\begin{align}
    Z^{l} = f_{\text{ENC}}^{l}(Z^{l-1}), 
\end{align}
where $Z^{l} \in \mathbb{R}^{T \times D}$ is the $l$-th layer output, with $Z^{0} = X$. 

Unlike MMS and XEUS, which directly process raw waveforms, OWSM-CTC~\cite{peng2024owsmctc} takes log Mel filter banks as input, which are processed by a convolutional encoder followed by $L$ stacked E-Branchformer~\cite{10022656ebranchformer} encoder layers ($\{f_{\text{ENC}}^l\}_{l=1}^L$).
OWSM-CTC employs an encoder-only architecture with a prompt encoder.
For fair comparison in ML-SUPERB~2.0, we evaluate only its speech encoder with speech input, excluding prefix tokens and other components. 

\subsubsection{Training downstream models with frozen upstream}
\label{sec:downstream}
To assess the cross-lingual speech representation ability of SSL and supervised model encoders, we use them as upstream models and stack Transformer encoder layers as downstream models. 
These downstream layers further model contextual dependencies given the upstream representations.
During training, the upstream parameters are frozen and only the downstream layers are optimized. 
To aggregate upstream representations, we apply a weighted sum approach~\cite{matthew2018featurizer,yang21s3prl} to derive the final output:
\begin{align}
    Z_{\text{out}} = \sum_{l = 1}^{L} \alpha_l \cdot Z^{l},
\end{align}
where $\alpha_l$ is a learnable parameter satisfying $\sum_{l = 1}^{L} \alpha_l = 1$.

To adapt upstream representations, $Z_{\text{out}} \in \mathbb{R}^{T\times D}$ is linearly projected to a lower feature dimension of $D_{\text{proj}}$ and then mapped to the downstream hidden dimension $D_{\text{down}}$ through an input layer using either a subsampling convolution (with factor $\kappa$) or a linear layer (i.e. $\kappa$ = 1). 
Positional encoding is then added before feeding the transformed features into the subsequent downstream encoder layers, producing the final output $H_{\text{out}} \in \mathbb{R}^{T / \kappa \times D_{\text{down}}}$. 

Given the $S$-length target sequence $Y = (y_s \in \mathcal{V} | s = 1, \dots, S)$, where $\mathcal{V}$ is a vocabulary consisting of linguistic tokens and language codes, the CTC loss is defined as:
\begin{align}
    \label{eq:ctc}
    P_{\text{CTC}}(Y|H_{\text{out}}) &= \sum_{\pi \in \mathcal{F}^{-1}(Y)} \prod_{t=1}^{T / \kappa} P(\pi_t | H_{\text{out}}), \\
    \mathcal{L}_{\text{CTC}} &= -\log P_{\text{CTC}}(Y|H_{\text{out}}),
\end{align}
where $\mathcal{F}^{-1}(Y)$ denotes the set of all latent alignments between the downstream output $H_{\text{out}}$ and target sequence $Y$, and $\pi$ represents a specific alignment. 
During training, we freeze the upstream and fully optimize the downstream using the CTC loss. 

\begin{figure}[t]
  \centering
  \includegraphics[width=\linewidth]{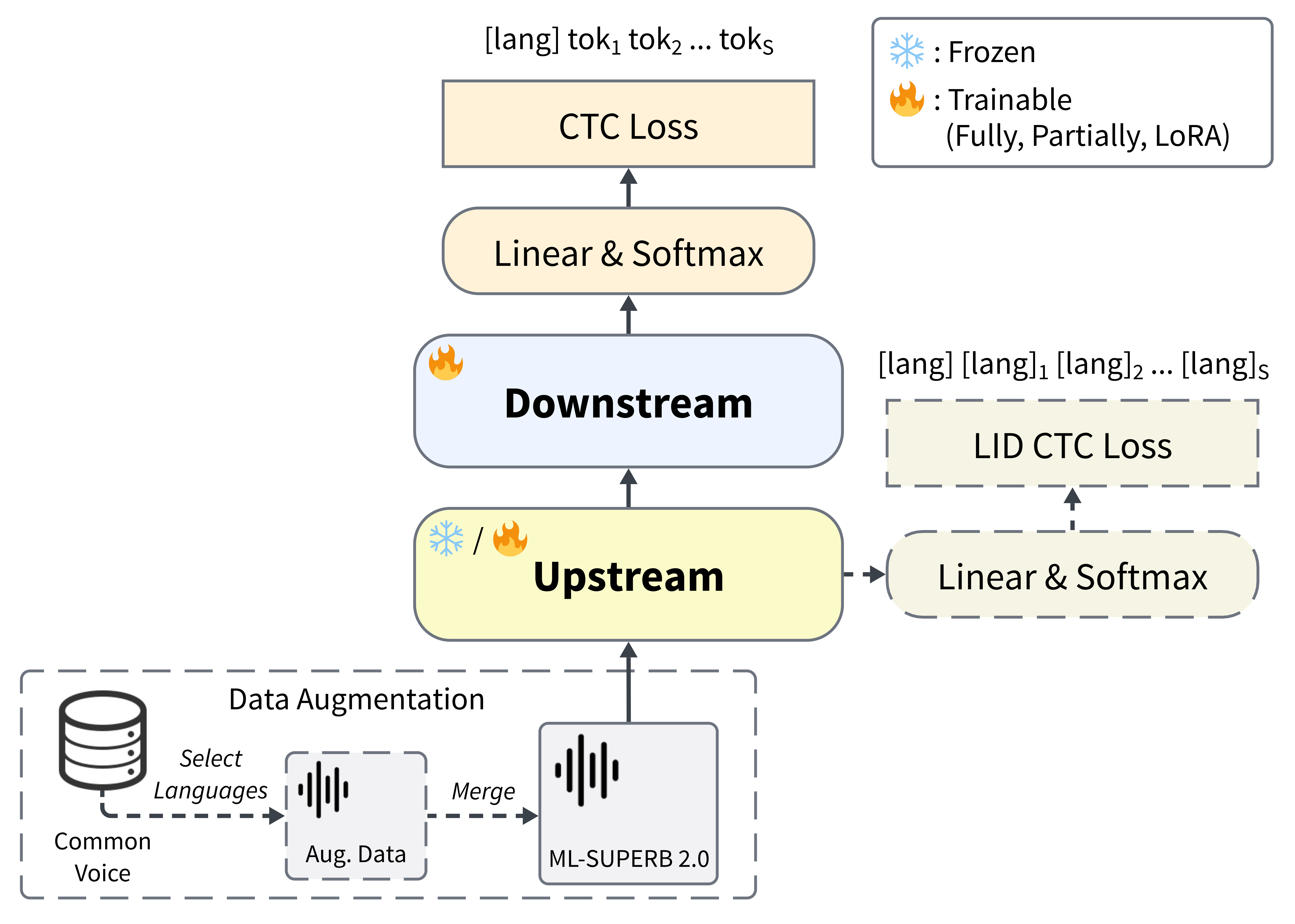}
  \caption{Model architecture. The upstream model is frozen, partially fine-tuned, or adapted with LoRA. The downstream model is fully trainable. Training uses CTC loss, with optional data augmentation and LID CTC loss. Dashed lines denote optional enhancements; solid lines represent core processes.}
  \label{fig:model}
\end{figure}

\subsubsection{Fine-tuning upstream models}
\label{sec:finetuning}
While training only the downstream layers leverages pre-trained representations from the upstream model, it may limit the upstream model's adaptability to target languages. 
To address this, we fine-tune the upstream model by unfreezing partial layers rather than all layers, as full-layer fine-tuning is computationally expensive. 
Specifically, we optimize the bottom, middle, or top consecutive layers $\mathcal{N} \subset \{1, \dots, L\}$, with the number and position of unfrozen layers varying across different upstream models. 
During fine-tuning, we use the same downstream architectures as in Section~\ref{sec:downstream}, but with shallower encoder layers, optimized with the CTC loss to better adapt upstream representations for multilingual LID and ASR. 

\subsubsection{Parameter-efficient upstream model adaptation}
\label{sec:lora}
Full-parameter fine-tuning is computationally expensive due to numerous trainable parameters.  
To optimize efficiently, we use LoRA~\cite{Hu2021LoRALA}, which freezes pre-trained parameters and injects trainable low-rank matrices into self-attention layers. 
We apply LoRA to all self-attention projection weights at each upstream layer. 
As before, shallow encoder layers are used in the downstream model, with the full model optimized with the CTC loss. 

\subsection{Data augmentation for low-resource languages}
\label{sec: Data Augmentation}
The ML-SUPERB~2.0 official training data includes 20 few-shot languages, each with at most 5 utterances in the training set~\cite{ml-superb2}. 
Due to limited resources, both LID and ASR performance degrade from insufficient language-specific supervision. 
To address this, we augment the ML-SUPERB~2.0 official training set with additional data for underperforming languages, as illustrated in Figure~\ref{fig:model}.
Specifically, we select data from the Common Voice Corpus~\cite{ardila2019common} (version 20.0), covering some few-shot and underperforming languages present in the baseline results.\footnote{ISO~639-3 codes of augmented languages: vie, srp, lit, dan, frr, tur, kaz, zul, epo, tok, ory, ast, lao, nan, cmn, kat, kor, jpn, yue, lit, kab, heb.}
For each language, we randomly sample an average of 2123 utterances, with exact counts varying. 
To mitigate the language bias, the sampled data is integrated into the ML-SUPERB~2.0 official training data, and the models are trained with the augmented training set.

\subsection{Auxiliary LID CTC loss}
\label{sec:lidctc}
ML-SUPERB 2.0 provides an average of 1.6 hours of training data per language, which can lead to overfitting during fine-tuning.
To regularize fine-tuning and improve model generalizability, we introduce an auxiliary LID prediction task with an LID CTC objective, as depicted in Figure~\ref{fig:model}. 
Specifically, we define a language code sequence $I = (i_s \in \mathcal{I} \mid s = 1, \dots, S)$, where $\mathcal{I}$ is the set of language codes corresponding to languages in the training set. 
Each $i_s$ in $I$ is identical, representing the same language for the entire utterance. 
The sequence length $S$ matches that of the transcription sequence $Y$ in Equation~\ref{eq:ctc}.
Given the output $Z^l$ of the $l$-th upstream fine-tuning encoder layer $f_{\text{ENC}}^l$, the LID CTC loss for layer $l$ is defined as: 
\begin{align}
    P_{\text{CTC}}^l(I | Z^l) &= \sum_{\varphi \in \mathcal{G}^{-1}(I)}\prod_{t=1}^{T}P(\varphi_t | Z^l), \\
    \mathcal{L}_{\text{LID}}^l &= -\log P_{\text{CTC}}^l(I | Z^l),
\end{align}
where $\mathcal{G}^{-1}(I)$ represents the set of possible alignments between $Z^l$ and $I$, and $\varphi$ denotes a specific alignment. 
The LID CTC loss is applied to a subset $\mathcal{M}$ of fine-tuning layers, where $\mathcal{M} \subseteq \mathcal{N}$ and $\mathcal{N}$ denotes the set of fine-tuning layers mentioned in Section~\ref{sec:finetuning}.
The final training objective combines the primary ASR CTC loss with the auxiliary LID CTC loss:
\begin{align}
    \label{eq:loss}
    \mathcal{L} = (1 - \beta) \mathcal{L}_{\text{CTC}} + \beta\frac{\sum_{l \in \mathcal{M}}\mathcal{L}_{\text{LID}}^l}{|\mathcal{M}|}, \quad \text{where } \beta \in [0, 1].
\end{align}

\section{Experiments}

\textbf{General setup.} 
We use the ML-SUPERB~2.0 official dataset from the ML-SUPERB Challenge 2025, which includes a training set (220 hours, 138 languages), standard development set (41 hours, 138 languages), and dialect development set (9 hours, 56 dialects).\footnote{\href{https://huggingface.co/datasets/espnet/ml_superb_hf}{https://huggingface.co/datasets/espnet/ml\_superb\_hf}} 
Models are evaluated on the development sets due to the unavailability of the test set.
We adopt the baseline provided by the challenge, using frozen MMS with \SI{1}{\billion} parameters (MMS-1B) as the frozen upstream model, with its layer outputs weighted summed and fed into a two-layer Transformer encoder.\footnote{\href{https://huggingface.co/espnet/mms_1b_mlsuperb}{https://huggingface.co/espnet/mms\_1b\_mlsuperb}}
We re-run this baseline model in our evaluation framework to ensure fair comparison, rather than using pre-reported results. 
We also report results based on XEUS~\cite{chen2024xeus} and XLS-R~128 (\SI{300}{\million})~\cite{babu2022xlsr} provided by the challenge, although the implementation details are not disclosed.
The largest feasible SFMs or newest variants are used as our upstream models: MMS-1B\footnote{\href{https://huggingface.co/facebook/mms-1b}{https://huggingface.co/facebook/mms-1b}} (\SI{1}{\billion} parameters), XEUS\footnote{\href{https://huggingface.co/espnet/xeus}{https://huggingface.co/espnet/xeus}} (\SI{577}{\million}), and OWSM-CTC v4 medium (\SI{1.01}{\billion})~\cite{anonymous2025owsmctcv4}.
A character tokenizer is used for all models.
All experiments are conducted using ESPnet~\cite{watanabe2018espnet}, with S3PRL~\cite{yang21s3prl} serving as the upstream model interface. 

\textbf{Model configuration.} 
Each model has projection dimension $D_{\text{proj}}$ = 80 (Section~\ref{sec:downstream}).
The input layer uses a subsampling convolutional neural network with factor $\kappa$ = 2 and kernel size 3, reducing the frame rate to \SI{40}{ms} for MMS and XEUS, while OWSM-CTC uses a linear layer to retain \SI{80}{ms}.
The downstream models employ Transformer encoders with sinusoidal positional encodings~\cite{vaswani2017transformer}. 
To ensure fair comparisons, configurations for frozen upstream training and upstream fine-tuning are designed to maintain a consistent trainable parameter size (around \SI{260}{\million}). 
The downstream model for training with frozen upstream consists of 12 layers, 1320 hidden units ($D_{\text{down}}$ in Section~\ref{sec:downstream}), 8 heads, and 5280 projection units. 
For upstream fine-tuning, it consists of 2 layers, 512 hidden units ($D_{\text{down}}$), 8 heads, and 2048 projection units. 
The specific fine-tuned upstream layers, selected from the bottom, middle, or top layers, are listed in Table~\ref{tab:all}, with the layer count adjusted to keep trainable parameters consistent.
The LoRA-based strategy follows the same setting but with 1 layer downstream encoder, and its rank and scaling factor are set to 16. 

\textbf{Training setup.} 
All models are trained for 10 epochs, with a batch size of 8 and 30000 steps in each epoch. 
Gradient accumulation is applied every 4 steps to stabilize updates. 
The learning rate is tuned over \numlist{1e-5;3e-5;1e-4;3e-4} with \num{1e-4} chosen, optimized with Adam optimizer~\cite{Kingma2014AdamAM}. 
A 25000-step warm-up is used for training downstream with frozen upstream and LoRA except for upstream fine-tuning. 
SpecAugment~\cite{park19e_interspeech_specaug} is applied to the upstream output. 
The final model is selected based on the lowest validation loss.

\textbf{Fine-tuning with augmented data and LID CTC.} 
Based on Table~\ref{tab:all}, we select the optimal fine-tuning layer set $\mathcal{N}$ for each upstream model: \{25$,\dots,$36\} for MMS-1B, \{12$,\dots,$19\} for XEUS, and \{8$,\dots,$13\} for OWSM-CTC. 
These configurations are then used for training on augmented data (Section~\ref{sec: Data Augmentation}).
For the LID CTC loss (Section~\ref{sec:lidctc}), the applied layer subset $\mathcal{M}$ is \{27$,$ 30$,$ 33$,$ 36\} for MMS-1B, \{14$,$ 17\} for XEUS, and \{10$,$ 13\} for OWSM-CTC.
We evaluate it on both the official and augmented ML-SUPERB~2.0 training sets, tuning $\beta$ in Equation~\ref{eq:loss} over \numlist{3e-2;1e-1;3e-1} with \num{3e-1} chosen.


\begin{table}[t]
    \caption{Results of different models and strategies on ML-SUPERB~2.0. LID accuracy (ACC, \%) and ASR CER (\%) are averaged across languages for full and worst-15 (WL) in the standard set, and dialects (DL). An asterisk (*) on Baseline denotes an unknown training strategy. T. indicates Transformer downstream. Numbers (e.g., 1-12) represent fine-tuned layers. Best per model is \underline{underlined}, overall best is \textbf{bolded}.}
    \label{tab:all}
    \centering
    {\fontsize{8pt}{9pt} \selectfont
    \renewcommand{\arraystretch}{1.2}
    \begin{tabular}{llccccc}
        \toprule
        \multirow{2.5}{*}{Model} & \multirow{2.5}{*}{Method} & \multicolumn{2}{c}{LID (ACC $\uparrow$)} & \multicolumn{3}{c}{ASR (CER $\downarrow$)} \\
        \cmidrule{3-4} \cmidrule(l){5-7}
        & & Full & DL & Full & WL & DL \\
        \midrule
        \multirow{1}{*}{MMS-1B} & Baseline & 76.1 & 58.4 & 22.3 $\pm$ 14.5 & 55.4 & 32.7 \\
        \multirow{1}{*}{XEUS} & Baseline* & 77.1 & 79.1 & 22.4 $\pm$ 30.9 & 78.9 & 23.2 \\
        \multirow{1}{*}{XLS-R 128} & Baseline* & 72.4 & 60.7 & 31.7 $\pm$ 26.9 & 86.8 & 30.5 \\
        \midrule
        \multirow{6}{*}{MMS-1B} 
        & T. & 69.6 & 50.6 & 27.9 $\pm$ 18.8 & 72.6 & 41.7 \\
        & 1-12 & 80.7 & 63.9 & 20.5 $\pm$ 14.7 & 54.5 & 38.2 \\
        & 13-24 & 81.5 & 68.4 & 18.9 $\pm$ 13.7 & 49.3 & 36.8  \\
        & 25-36 & 80.2 & \underline{\textbf{68.4}} & 17.1 $\pm$ 13.0 & 46.6 & 34.5 \\
        & 37-48 & 78.2 & 52.3 & 20.7 $\pm$ 15.5 & 55.3 & 43.9 \\
        & LoRA & \underline{\textbf{82.4}} & 64.9 & \underline{\textbf{16.3 $\pm$ 12.7}} & \underline{\textbf{44.5}} & \underline{\textbf{28.5}} \\
        \midrule
        \multirow{5}{*}{XEUS} 
        & T. & \underline{79.2} & 61.5 & 23.8 $\pm$ 12.8 & \underline{50.2} & 39.9\\
        & 1-8 & 74.7 & 52.7 & 25.6 $\pm$ 14.7 & 57.5 & 47.6\\
        & 7-14 & 77.0 & \underline{67.9} & 23.6 $\pm$ 14.7 & 57.0 & \underline{38.6}\\
        & 12-19 & 77.3 & 56.9 & \underline{21.6 $\pm$ 14.8} & 54.9 & 43.1\\
        & LoRA & 76.6 & 48.8 & 26.5 $\pm$ 18.2 & 69.4 & 43.9\\
        \midrule
        \multirow{6}{*}{OWSM-CTC} 
        & T. & 57.9 & 38.4 & 45.8 $\pm$ 18.3 & 83.2 & 66.4 \\
        & 1-6 & \underline{79.0} & 66.7 & 30.2 $\pm$ 15.7 & 63.6 & 50.1 \\
        & 8-13 & 78.3 & \underline{67.7} & \underline{28.2 $\pm$ 13.5} & \underline{56.4} & \underline{49.9} \\
        & 15-20 & 74.3 & 62.1 & 29.5 $\pm$ 13.5 & 57.7 & 52.7 \\
        & 22-27 & 63.4 & 52.5 & 37.1 $\pm$ 19.5 & 80.8 & 60.6 \\
        & LoRA & 74.3 & 64.3 & 32.4 $\pm$ 13.0 & 59.7 & 52.8 \\
        \bottomrule
    \end{tabular}
    }
\end{table}

\section{Results}

\subsection{Training strategies with multilingual SFMs}

Table~\ref{tab:all} shows LID and ASR results across SFMs and training strategies on ML-SUPERB~2.0. 
On the standard set, MMS-1B with LoRA achieves the best performance (82.4\% LID accuracy, 16.3\% CER).
On the dialect set, fine-tuning the middle layers of MMS-1B achieves the highest LID accuracy (68.4\%), while LoRA on MMS-1B yields the lowest CER (28.5\%).
XEUS performs competitively but lags behind MMS-1B, likely due to its smaller size (about 60\% of MMS-1B). 
Notably, our best strategies with MMS-1B and XEUS outperform all baselines on the standard set but trail the XEUS baseline on the dialect set.
Additionally, our methods exhibit lower overall deviations, suggesting improved robustness across languages. 
OWSM-CTC performs the worst, especially in ASR, as its pre-training lacks full ML-SUPERB~2.0 language coverage. 

\textbf{Which SFM is more effective for downstream?} 
In our setting, XEUS outperforms MMS-1B and OWSM-CTC when training with frozen upstream representations. 
Compared to the MMS-1B (Baseline), MMS-1B (T.) with a larger downstream model degrades performance, likely due to overfitting from increased depth and hidden size.
OWSM-CTC (T.) underperforms the most, likely due to the omission of the prefix token input~\cite{peng2024owsmctc} and its restricted language coverage. 

\textbf{Fine-tune which layers is more effective?}
For MMS-1B, fine-tuning middle layers (13-24, 25-36) outperforming bottom (1-12) and top layers (37-48), especially in dialect LID and standard ASR.
However, XEUS benefits more from fine-tuning upper layers on the standard set, achieving its best fine-tuning performance on 12-19 layers, while for dialect LID and ASR, mid-level layers (7-14) perform best.
Similar to MMS-1B, OWSM-CTC achieves lower CER by fine-tuning middle layers (8-13, 15-20) than bottom or top layers.
These results indicate that different models exhibit varying layer-wise encoding properties, requiring model-specific fine-tuning strategies.

\textbf{LoRA benefits MMS-1B more.}
The Transformer-based MMS-1B with LoRA achieves the best overall performance over fine-tuning and frozen upstream training.
However, LoRA for E-Branchformer-based XEUS and OWSM-CTC fails to surpass fine-tuning.
This may be because our LoRA configuration adapts only self-attention projections, which are central to Transformer.
In contrast, E-Branchformer's hybrid design with a local extractor~\cite{10022656ebranchformer} may require broader adaptation, limiting LoRA's effectiveness.

\begin{table}[t]
    \caption{Results on fine-tuning with data augmentation and LID CTC. Metrics follow Table~\ref{tab:all}, with results for few-shot (FS) languages in the standard set (Std.) explicitly reported. Numbers indicate fine-tuned layers. LIDCTC: fine-tuning with LID CTC on official data. DataAug: training on augmented data. DataAug*: fine-tuning with LID CTC on augmented data.}
    \label{tab:2}
    \centering
    {\fontsize{8pt}{9pt} \selectfont
    \renewcommand{\arraystretch}{1.2}
    \begin{tabular}{llccccccccc}
        \toprule
        \multirow{4}{*}{Model} & \multirow{4}{*}{Method} & \multicolumn{3}{c}{LID (ACC $\uparrow$)} & \multicolumn{4}{c}{ASR (CER $\downarrow$)} \\
        \cmidrule{3-5} \cmidrule(l){6-9}
        & & \multicolumn{2}{c}{Std.} & \multirow{2.5}{*}{DL} & \multicolumn{3}{c}{Std.} & \multirow{2.5}{*}{DL} \\
        \cmidrule{3-4} \cmidrule(l){6-8}
        & & Full & FS &  & Full & FS & WL & \\
        \midrule 
        \multirow{4}{*}{MMS-1B} & 25-36 & 80.2 & 2.4 & 68.4 & 17.1 $\pm$ 13.0 & 24.6 & 46.6 & 34.5 \\
        & LIDCTC & 82.9 & 6.1 & 66.2 & 17.1 $\pm$ 13.5 & 27.9 & 47.7 & 30.3 \\
        & DataAug & 86.0 & 40.0 & 63.9 & 15.9 $\pm$ 10.7 & 15.2 & 39.3 & 34.3 \\
        & DataAug* & 86.9 & 40.7 & 74.2 & 15.6 $\pm$ 10.5 & 15.1 & 38.6 & 31.5 \\
        \midrule
        \multirow{4}{*}{XEUS} & 12-19 & 77.3 & 0.0 & 56.9 & 21.6 $\pm$ 14.8 & 33.1 & 54.9 & 43.1\\
        & LIDCTC & 79.9 & 4.4 & 68.7 & 21.4 $\pm$ 14.6 & 33.1 & 55.4 & 34.8  \\
        & DataAug & 84.8 & 42.7 & 55.5 & 18.1 $\pm$ 10.5 & 15.9 & 41.5 & 39.4   \\
        & DataAug* & 85.8 & 44.6 & 69.5 & 17.9 $\pm$ 10.5 & 16.4 & 42.0 & 35.4 \\
        \midrule
        \multirow{4}{*}{OWSM-CTC} & 8-13 & 78.3 & 4.7 & 67.7 & 28.2 $\pm$ 13.5 & 33.8 & 56.4 & 49.9 \\
        & LIDCTC & 76.4 & 2.2 & 62.4 & 29.3 $\pm$ 13.2 & 33.3 & 57.0 & 52.2  \\
        & DataAug & 81.4 & 42.3 & 64.7 & 27.6 $\pm$ 11.7 & 21.7 & 50.5 & 52.4 \\
        & DataAug* &  82.0 & 39.9 & 64.5 & 27.5 $\pm$ 11.5 & 21.7 & 50.3 & 50.8  \\
        \bottomrule
    \end{tabular}
    }
\end{table}

\subsection{Fine-tuning with data augmentation and LID CTC}

\textbf{Data augmentation boosts LID and ASR on few-shot languages.}
Table~\ref{tab:2} presents the LID and ASR results for fine-tuning with data augmentation and LID CTC, explicitly considering performance on few-shot languages.
With data augmentation, the most significant improvements are observed in LID and ASR for few-shot languages, highlighting the benefit of additional training data for low-resource languages. 
For instance, fine-tuning XEUS (12-19) on augmented data improves few-shot language LID accuracy by 42.7\% and reduces CER by 17.2\% absolutely compared to using the official dataset. 
Similarly, MMS-1B and OWSM-CTC see notable performance improvements on few-shot languages, confirming the benefits of augmentation across various architectures.
This improvement suggests that data augmentation enhances model robustness by increasing data diversity and mitigating overfitting. 

\textbf{LID CTC enhances dialect performance for MMS-1B and XEUS. } 
Beyond data augmentation, applying LID CTC during fine-tuning introduces additional improvements for MMS-1B and XEUS, especially for dialects.
For XEUS fine-tuned on the official training set, dialect LID accuracy increases by 11.8\% and ASR CER decreases by 8.3\% absolutely.
While MMS-1B sees a slight drop in dialect LID accuracy on the official dataset, it benefits significantly from LID CTC when trained on augmented data, improving dialect LID accuracy by 10.3\% and reducing dialect ASR CER by 2.8\% absolutely. 
This suggests that auxiliary LID prediction enhances generalization, though sufficient data may be required.
Unlike MMS-1B and XEUS, applying LID CTC loss to OWSM-CTC leads to slight degradation with both official and augmented data. 
This suggests that the auxiliary LID CTC loss may interfere with OWSM-CTC's supervised pre-training, which is already optimized for LID and ASR, limiting further improvements. 

\section{Conclusion}

In this paper, we explore strategies to improve multilingual LID and ASR performance on ML-SUPERB~2.0. 
We evaluate MMS, XEUS, and OWSM-CTC under downstream training with frozen upstream, upstream fine-tuning, and LoRA. 
Furthermore, we incorporate data augmentation and LID CTC loss for fine-tuning. 
Results show that MMS-1B outperforms XEUS and OWSM-CTC under our strategies, especially with LoRA and mid-layer fine-tuning. 
Data augmentation improves few-shot performance, while LID CTC loss enhances generalization on dialects. 
Overall, our MMS-1B-based model, fine-tuned with LID CTC loss on augmented data, achieves \textbf{second place} in the Interspeech 2025 ML-SUPERB~2.0 Challenge. 
Future work will explore language model integration and strategies to further improve low-resource language performance.

\section{Acknowledgements}

Experiments of this work used the Bridges2 system at PSC and Delta system at NCSA through allocations CIS210014 and IRI120008P from the Advanced Cyberinfrastructure Coordination Ecosystem: Services \& Support (ACCESS) program, supported by National Science Foundation grants \#2138259, \#2138286, \#2138307, \#2137603, and \#2138296.

\bibliographystyle{IEEEtran}
\bibliography{mybib}  

\begin{thebibliography}{10}
\providecommand{\url}[1]{#1}
\csname url@samestyle\endcsname
\providecommand{\newblock}{\relax}
\providecommand{\bibinfo}[2]{#2}
\providecommand{\BIBentrySTDinterwordspacing}{\spaceskip=0pt\relax}
\providecommand{\BIBentryALTinterwordstretchfactor}{4}
\providecommand{\BIBentryALTinterwordspacing}{\spaceskip=\fontdimen2\font plus
\BIBentryALTinterwordstretchfactor\fontdimen3\font minus \fontdimen4\font\relax}
\providecommand{\BIBforeignlanguage}[2]{{%
\expandafter\ifx\csname l@#1\endcsname\relax
\typeout{** WARNING: IEEEtran.bst: No hyphenation pattern has been}%
\typeout{** loaded for the language `#1'. Using the pattern for}%
\typeout{** the default language instead.}%
\else
\language=\csname l@#1\endcsname
\fi
#2}}
\providecommand{\BIBdecl}{\relax}
\BIBdecl

\bibitem{austin2011camblang}
J.~S. Peter K.~Austin, Ed., \emph{The Cambridge Handbook of Endangered Languages}, ser. Cambridge Handbooks in Language and Linguistics.\hskip 1em plus 0.5em minus 0.4em\relax Cambridge University Press, 2011.

\bibitem{babu2021xls}
\BIBentryALTinterwordspacing
A.~Babu, C.~Wang, A.~Tjandra, K.~Lakhotia \emph{et~al.}, ``{XLS-R: Self-supervised Cross-lingual Speech Representation Learning at Scale},'' in \emph{Proc. Interspeech}, 2021, pp. 2278--2282.
\BIBentrySTDinterwordspacing

\bibitem{li2022asr2k}
X.~Li, F.~Metze, D.~R. Mortensen, A.~W. Black \emph{et~al.}, ``{ASR2K}: Speech recognition for around 2000 languages without audio,'' in \emph{Proc. Interspeech}, 2022, pp. 4885--4889.

\bibitem{chen2023wavlablm}
W.~Chen, J.~Shi, B.~Yan, D.~Berrebbi \emph{et~al.}, ``Joint prediction and denoising for large-scale multilingual self-supervised learning,'' in \emph{Proc. ASRU}, 2023, pp. 1--8.

\bibitem{pratap2024mms}
V.~Pratap, A.~Tjandra, B.~Shi, P.~Tomasello \emph{et~al.}, ``Scaling speech technology to 1,000+ languages,'' \emph{Journal of Machine Learning Research}, vol.~25, no.~97, pp. 1--52, 2024.

\bibitem{chen2024xeus}
\BIBentryALTinterwordspacing
W.~Chen, W.~Zhang, Y.~Peng, X.~Li \emph{et~al.}, ``Towards robust speech representation learning for thousands of languages,'' in \emph{Proc. EMNLP}, 2024, pp. 10\,205--10\,224.
\BIBentrySTDinterwordspacing

\bibitem{boito2024mhubert}
M.~{Zanon Boito}, V.~Iyer, N.~Lagos, L.~Besacier \emph{et~al.}, ``{mHuBERT-147}: A compact multilingual {HuBERT} model,'' in \emph{Proc. Interspeech}, 2024, pp. 3939--3943.

\bibitem{zhou2021acm}
\BIBentryALTinterwordspacing
L.~Zhou, J.~Li, E.~Sun, and S.~Liu, ``A configurable multilingual model is all you need to recognize all languages,'' in \emph{Proc. ICASSP}, 2021, pp. 6422--6426.
\BIBentrySTDinterwordspacing

\bibitem{radford2023whisper}
A.~Radford, J.~W. Kim, T.~Xu, G.~Brockman \emph{et~al.}, ``Robust speech recognition via large-scale weak supervision,'' in \emph{Proc. ICML}, 2023, pp. 28\,492--28\,518.

\bibitem{peng2024owsm}
Y.~Peng, J.~Tian, W.~Chen, S.~Arora \emph{et~al.}, ``{OWSM} v3.1: Better and faster open {Whisper}-style speech models based on {E-Branchformer},'' in \emph{Proc. Interspeech}, 2024, pp. 352--356.

\bibitem{peng2024owsmctc}
Y.~Peng, Y.~Sudo, M.~Shakeel, and S.~Watanabe, ``{OWSM-CTC}: An open encoder-only speech foundation model for speech recognition, translation, and language identification,'' in \emph{Proc. ACL}, 2024, pp. 10\,192--10\,209.

\bibitem{ml-superb2}
J.~Shi, S.-H. Wang, W.~Chen, M.~Bartelds \emph{et~al.}, ``{ML-SUPERB 2.0}: Benchmarking multilingual speech models across modeling constraints, languages, and datasets,'' in \emph{Proc. Interspeech}, 2024, pp. 1230--1234.

\bibitem{ml-superb}
J.~Shi, D.~Berrebbi, W.~Chen, H.-L. Chung \emph{et~al.}, ``{ML-SUPERB}: Multilingual speech universal performance benchmark,'' in \emph{Proc. Interspeech}, 2023, pp. 884--888.

\bibitem{graves2006ctc}
A.~Graves, S.~Fern{\'a}ndez, F.~Gomez, and J.~Schmidhuber, ``Connectionist temporal classification: labelling unsegmented sequence data with recurrent neural networks,'' in \emph{Proc. ICML}, 2006, pp. 369--376.

\bibitem{tjandra2020deja}
A.~Tjandra, C.~Liu, F.~Zhang, X.~Zhang \emph{et~al.}, ``Deja-vu: Double feature presentation and iterated loss in deep transformer networks,'' in \emph{Proc. ICASSP}, 2020, pp. 6899--6903.

\bibitem{lee2021intermediate}
J.~Lee and S.~Watanabe, ``Intermediate loss regularization for {CTC}-based speech recognition,'' in \emph{Proc. ICASSP}, 2021, pp. 6224--6228.

\bibitem{nozaki2021relaxing}
J.~Nozaki and T.~Komatsu, ``Relaxing the conditional independence assumption of {CTC}-based {ASR} by conditioning on intermediate predictions,'' in \emph{Proc. Interspeech}, 2021, pp. 3735--3739.

\bibitem{chen2023lidctc}
W.~Chen, B.~Yan, J.~Shi, Y.~Peng \emph{et~al.}, ``Improving massively multilingual {ASR} with auxiliary {CTC} objectives,'' in \emph{Proc. ICASSP}, 2023, pp. 1--5.

\bibitem{li2022lifelong}
B.~Li, R.~Pang, Y.~Zhang, T.~N. Sainath \emph{et~al.}, ``Massively multilingual {ASR}: A lifelong learning solution,'' in \emph{Proc. ICASSP}, 2022, pp. 6397--6401.

\bibitem{barteldsetal2023making}
\BIBentryALTinterwordspacing
M.~Bartelds, N.~San, B.~McDonnell, D.~Jurafsky \emph{et~al.}, ``Making more of little data: Improving low-resource automatic speech recognition using data augmentation,'' in \emph{Proc. ACL}, 2023, pp. 715--729.
\BIBentrySTDinterwordspacing

\bibitem{ko2015speedperturb}
\BIBentryALTinterwordspacing
T.~Ko, V.~Peddinti, D.~Povey, and S.~Khudanpur, ``Audio augmentation for speech recognition,'' in \emph{Proc. Interspeech}, 2015, pp. 3586--3589.
\BIBentrySTDinterwordspacing

\bibitem{park19e_interspeech_specaug}
D.~S. Park, W.~Chan, Y.~Zhang, C.-C. Chiu \emph{et~al.}, ``Specaugment: A simple data augmentation method for automatic speech recognition,'' in \emph{Proc. Interspeech}, 2019, pp. 2613--2617.

\bibitem{xue2024sshr}
H.~Xue, Q.~Shao, K.~Huang, P.~Chen \emph{et~al.}, ``{SSHR}: Leveraging self-supervised hierarchical representations for multilingual automatic speech recognition,'' in \emph{Proc. ICME}, 2024, pp. 1--6.

\bibitem{vaswani2017transformer}
A.~Vaswani, N.~Shazeer, N.~Parmar, J.~Uszkoreit \emph{et~al.}, ``Attention is all you need,'' in \emph{Proc. NeurIPS}, 2017, pp. 6000--6010.

\bibitem{10022656ebranchformer}
K.~Kim, F.~Wu, Y.~Peng, J.~Pan \emph{et~al.}, ``{E-Branchformer}: Branchformer with enhanced merging for speech recognition,'' in \emph{Proc. SLT}, 2023, pp. 84--91.

\bibitem{matthew2018featurizer}
M.~E. Peters, M.~Neumann, M.~Iyyer, M.~Gardner \emph{et~al.}, ``Deep contextualized word representations,'' in \emph{Proc. NAACL}, 2018, pp. 2227--2237.

\bibitem{yang21s3prl}
S.~wen Yang, P.-H. Chi, Y.-S. Chuang, C.-I.~J. Lai \emph{et~al.}, ``{SUPERB}: Speech processing universal performance benchmark,'' in \emph{Proc. Interspeech}, 2021, pp. 1194--1198.

\bibitem{Hu2021LoRALA}
\BIBentryALTinterwordspacing
E.~J. Hu, yelong shen, P.~Wallis, Z.~Allen-Zhu \emph{et~al.}, ``Lo{RA}: Low-rank adaptation of large language models,'' in \emph{Proc. ICLR}, 2022.
\BIBentrySTDinterwordspacing

\bibitem{ardila2019common}
\BIBentryALTinterwordspacing
R.~Ardila, M.~Branson, K.~Davis, M.~Kohler \emph{et~al.}, ``Common voice: A massively-multilingual speech corpus,'' in \emph{Proc. LREC}, 2020, pp. 4218--4222.
\BIBentrySTDinterwordspacing

\bibitem{babu2022xlsr}
A.~Babu, C.~Wang, A.~Tjandra, K.~Lakhotia \emph{et~al.}, ``{XLS-R}: Self-supervised cross-lingual speech representation learning at scale,'' in \emph{Proc. Interspeech}, 2022, pp. 2278--2282.

\bibitem{anonymous2025owsmctcv4}
Y.~Peng, S.~Muhammad, Y.~Sudo, W.~Chen \emph{et~al.}, ``{OWSM v4}: Improving open {Whisper}-style speech models via data scaling and cleaning,'' in \emph{Proc. Interspeech}, 2025.

\bibitem{watanabe2018espnet}
\BIBentryALTinterwordspacing
S.~Watanabe, T.~Hori, S.~Karita, T.~Hayashi \emph{et~al.}, ``{ESPnet}: End-to-end speech processing toolkit,'' in \emph{Proc. Interspeech}, 2018, pp. 2207--2211.
\BIBentrySTDinterwordspacing

\bibitem{Kingma2014AdamAM}
\BIBentryALTinterwordspacing
D.~P. Kingma and J.~Ba, ``Adam: {A} method for stochastic optimization,'' in \emph{Proc. ICLR (Poster)}, 2015.
\BIBentrySTDinterwordspacing

\end{thebibliography}
\end{document}